\documentclass[epj,referee]{svjour}
\usepackage{graphics}
\usepackage{slashbox}
\begin{document}
\title{The effect of the forget-remember mechanism on spreading}
\author{Gu J.\inst{1} \and Li W.\inst{1,2,3}
\thanks{To whom correspondence should be addressed. Electronic address: liwei@mis.mpg.de.} \and Cai X.\inst{1}% etc
}                     % Do not remove
\institute{Complexity Science Center, and Institute of Particle
Physics, Hua-Zhong Normal University, Wuhan 430079, China \and
Max-Planck-Institute for Mathematics in the Sciences, Inselstr. 22,
04103 Leipzig, Germany \and Santa Fe Institute, 1399 Hyde Park Rd,
NM 87501, USA }
\date{Received: date / Revised version: date}
\abstract{We introduce a new mechanism---the forget-remember
mechanism into the spreading process. Equipped with such a mechanism
an individual is prone to forget the ``message" received and
remember the one forgotten, namely switching his state between
active (with message) and inactive (without message). The
probability of state switch is governed by linear or exponential
forget-remember functions of history time which is measured by the
time elapsed since the most recent state change. Our extensive
simulations reveal that the forget-remember mechanism has
significant effects on the saturation of message spreading, and may
even lead to a termination of spreading under certain conditions.
This finding may shed some light on how to control the spreading of
epidemics. It is found that percolation-like phase transitions can
occur. By investigating the properties of clusters, formed by
connected, active individuals, we may be able to justify the
existence of such phase transitions.
\PACS{
      {89.75.Fb}{Structures and organization in complex systems}   \and
      {89.70.+c}{Information theory and communication theory}   \and
      {89.75.Hc}{Networks and genealogical trees}
     } % end of PACS codes
} %end of abstract
\maketitle
\section{Introduction}
\label{intro} The spreading process, through which news, rumors and
diseases, etc., can be transmitted, is ubiquitous in nature
\cite{Refz1,Refz2,Refz3}. Recent research along this topic has been
largely focused on the modelling of epidemics
\cite{Refz4,Refz5,Refz6,Refz7,Refz8,Refz9,Refz10} and its interplay
with biological interactions, which has yielded many valuable and
interesting results \cite{Refz11,Refz12}. Some of these models
studied have attracted the attention of epidemiologists
\cite{Refz13,Refz14,Refz15,Refz16,Refz17,Refz18}. Furthermore, it
has been pointed out that epidemiological processes can be related
to the well-known percolation
\cite{Refz8,Refz19,Refz20,Refz21,Refz22,Refz23}. In some other
models \cite{Refz24,Refz25,Refz26,Refz27,Refz28,Refz29}, the
epidemic spreading has been fully analyzed on different types of
networks to see its dependence on spatial effects.

In this paper, we employ the general term ``$\bf message$" to
refer to any object that can be transmitted in various spreading
processes. Henceforth, in this sense, types of messages are very
diverse, from computer viruses, e-mail, rumors, to forest fires
and contagious diseases (such as flu) and so on and so forth
\cite{Refz13,Refz30}. Most spreading processes share the following
common features: (i) Messages may not only be spread, but be
``$\bf forgotten$" and ``$\bf remembered$." In the whole context
of this paper ``forgotten" and ``remembered" are also general
terms. If the message means disease, then ``forgotten" is
equivalent to ``recovered" and ``remembered," ``reinfected." (ii)
Each member within a message-spreading system could be at either
of the two states, having a message or not. If he has a message,
an individual is opted to transmit it or not; otherwise, he can
accept or decline a message from others. Altogether there can be
four possible states for each individual, which are not totally
included by most models. For example, the SIS model for epidemics
assumes that each individual is either susceptible or infective
\cite{Refz31}. The SIR model adds a third one
\cite{Refz32,Refz33}---the removed state (the message is lost and
the individual never accepts the message ever again). Take the
smallpox spreading as an example \cite{Refz34}. A healthy human
being who never got this disease is in the susceptible state. His
state will turn into infective once he is infected by the disease
for the very first time. Right after his recovery, he will never
be infected by the smallpox again since he has acquired the
immunity against it. From this example we see clearly the state
transition from susceptible to infective and to removed. (iii)
Normally the spreading rate, which determines how quickly a
specific message can be transmitted, is quite limited.

Here we propose the forget-remember mechanism, realized by
respective probability functions which measure in a quantitative
way how the message can be forgotten and remembered, to study the
message spreading in a 2-state model. One state is active (with
message) and another, inactive (without message). We can simply
use ``1" to represent the active state and ``0," the inactive one.
We will focus on the effects of the forget-remember mechanism
(FRM) on the efficiency of message spreading. Namely, under what
conditions can a message be spread to all (or most of ) the
members of the population? By varying the parameters of the
forget- and remember function, are we able to prevent a message
from spreading at its infancy? If yes, what can we learn from it?
The answers of the above questions are important and relevant in
studying the breakout and the control of epidemics.

%\begin{figure*}
%\includegraphics{fig1.eps}
%\vspace*{1cm}       % Give the correct figure height in cm
%\caption{Probability functions for the forget-remember mechanism.
%$P_{-}(t)$ is the probability for the forget mechanism while
%$P_{+}(t)$ is the one for the remember mechanism, with both being
%ranged between 0 and 1. The left two panels correspond to the linear
%form, and the right two panels, the exponential one.}
%\label{fig.1}       % Give a unique label
%\end{figure*}

\section{The forget-remember mechanism and the model}

In most previous studies of spreading processes the most important
parameter---the effective spreading rate, determines not only the
percentage of active individuals, but whether a message will quickly
become popular like an epidemic \cite{Refz32,Refz33}. Here we take a
different way by relating the message spreading to the learning
process. In the message spreading, individuals forget and remember
messages as time elapses. For example, humans infected with diseases
like the flu, can recover even without taking any medication, but
the disease can also reoccur after a certain period of time.
Normally, the longer a person holds a message, the greater the
probability he will lose it, and the less the probability he will
remember it after a longer time. This feature also applies to the
well-known learning curve discovered by German experimental
psychologist Hermann Ebbinghaus \cite{Refz35,Refz36}. The
similarities between the learning curve and  the FRM suggest that
the former could be a guide to understanding the latter.

Our FRM is described as follows.

(i) Forget mechanism---when he holds a message, an individual may
forget it with probability $P_{-}(t)$, a function of history time
$t$. We assume that the longer a message is held, the more easily it
will be forgotten by its owner.

(ii) Remember mechanism---a forgotten message can be remembered,
with probability $P_{+}(t)$, also a function of time $t$.

(iii) The forget mechanism can be independent of the remember
mechanism, but the latter must rely on the former. Here previous
history of states counts in that an individual who never experienced
a forget-process would not remember any message.

(iv) The history time $t$ appeared in both $P_{-}(t)$ and
$P_{+}(t)$ is different from the system time $T$, because the
former is directly related to individual's previous states (the
notions of $T$ and $t$ are universal in the whole text without any
specific explanation). $t$ is defined as the time elapsed since
the individual's most recent state switch. Hence for the forget
mechanism, $t$ starts counting when a message is received; while
for the remember mechanism, it starts when a message is forgotten.

In order to gain further insight into the FRM, we consider the
linear form for both $P_{-}(t)$ and $P_{+}(t)$ (Eq. 1) with
parameters $a$ and $b$, and the exponential one for them, with
parameters $\alpha$ and $\beta$.

The linear function is simply
\begin{equation}
\label{eq.1} P_{\mp}(t)=a\pm bt %\right.
\end{equation}
and the exponential one is
\begin{equation}
\label{eq.1} P_{\mp}(t)=\alpha\mp e^{-\beta t}
%\right.
\end{equation}
\noindent In our simulations, parameters $b$ and $\beta$ are
chosen to be no less than zero. The probability functions must
take values between 0 and 1, so the range of the parameters is
accommodated accordingly (please refer to Fig. 1 for more
details). When both $b$ and $\beta$ are equal to zero, the
probability functions become uniform distributions. $a$ and
$\alpha\mp1$ are the initial values of the probability functions,
which could represent the importance of the message in the rumor
spreading process, or the initial probability of self-cure and
relapse in epidemics. $b$ and $\beta$, which determine the shapes
of two functions, can be regarded as the forget- and remember
speed. Namely, $b$ and $\beta$ show how quickly a message can be
forgotten and/or remembered.

Table 1, which shows the main correspondence of the standard SIS
and SIR with the corresponding message spreading forget-remember
mechanism, may be utilized to understand the FRM.

\begin{table*}
\caption{The main correspondence of the standard SIS and SIR with
the corresponding message spreading forget-remember mechanism}
\label{tab:1}       % Give a unique label
\begin{tabular}{|r|c|c|c|}
\hline\noalign{\smallskip}
\backslashbox[32mm]{Items}{Models}& SIS model & SIR model & our model with the FRM  \\
\noalign{\smallskip}\hline\noalign{\smallskip}
Number of individual's states & 2  & 3  & 2 \\
Probability of states' switch & constants & constants &
variable via probability functions\\
Means of message spreading & via neighbors & via neighbors & via
neighbors or remember mechanism\\
Means of message losing & self-recovery &
self-recovery & via forget mechanism\\
Transformation to SIS model & / & without the removed state & constant forget probability\\
Transformation to SIR model & including the removed state & / & no
remember mechanism\\
 \noalign{\smallskip}\hline
\end{tabular}
% Or use
\vspace*{5cm}  % with the correct table height
\end{table*}

Our model which incorporates the FRM is based on a scale-free
network, to which most social networks, including spreading
networks, belong. The degree distribution of the standard
Barab\'{a}si-Albert (BA) scale-free network is a power-law,
$P(k)\sim k^{-\gamma}$, with exponent $\gamma$ ranged between 2 and
3 \cite{Refz30,Refz37}. Therefore we build a standard BA scale-free
network of $N=10000$ with average degree $\langle k\rangle=4$ and
exponent $\gamma=2.7$.

Now we address the issue of how a message can be spread in our
model. The initial condition is that each individual in the
network has the equal chance to be ``infected" by an in-coming
message, with a very small probability $P_{\rm A}$. For instance,
if $P_{\rm A}$ is taken to be 0.005, then around 50 nodes are
initially activated and the rest ones remain inactive. Choosing
this tiny probability is reasonable when one takes the epidemics
as an example: at the infancy of an epidemic only a very small
fraction of the population is infected. Furthermore, to initialize
the spreading we assume that an individual who is inactive can be
activated by his active neighbors with transmission probability
$\nu$ for each, which is set to be very small in most of our
simulations. One might think that the interaction is very weak due
to $\nu$ being small. This is not always true when one considers
the way that each individual is connected. In a scale-free
network, the hubs may have several hundred nearest neighbors, so
they can be more easily infected, which enhances the chance for a
message to spread further away. This also makes sense in the case
of an epidemic, where the population size is of the magnitude of
million. In the absence of the FRM, the message will diffuse to
the whole system in a very prompt way. Now we introduce the forget
mechanism, i.e., an individual who is active may change his state
into inactive according to the probability function $P_{-}(t)$. If
there is only forget mechanism, then the message may stop
spreading eventually or does not spread in an efficient manner. So
we need to incorporate the remember mechanism, characterized by
$P_{+}(t)$. The value of $P_{+}(t)$ gives the probability that at
time $t$ an inactive individual changes his state to active. We
shall bear in mind that the activation of an inactive node is
co-determined by the interaction ($\nu$) and the remember
mechanism ($P_{+}(t)$). This is why we need to have small $\nu$
since the effects of the remember mechanism might be covered at
higher $\nu$.

We use $S_i(T)$ to denote the state of individual $i$ at a give time
$T$. According to our definition, $S_i(T)$ can only take two
distinct values, either 1 or 0. Due to the existence of the FRM, the
states evolution of the whole system is complex. In order for one to
get to know our model more clearly, let us follow the state change
of individual $i$ at any given time $T$.
\begin{enumerate}

\item If the state of $i$ at $T$ is $S_i(T)=1$, then $i$ changes his
state to $S_i(T+1)=0$ with probability $P_{-}(T-T^i_{0,1})$, where
$T^i_{0,1}$ is the most recent time when $i$ changes his state from
0 to 1. Now two consequences: $S_i(T+1)=0$ or $S_i(T+1)=1$. If the
former holds, then he starts to remember the message at time $T+1$,
or equivalently his remember time $t$ starts counting at $T+1$;
otherwise he still remains active at time $T+1$ but his forget time
is extended by 1.

\item If the state of $i$ at $T$ is $S_i(T)=0$, $i$ will calculate
the total number of its active nearest neighbors. If that number is
$A_i(T)$, then $i$ changes his states to $S_i(T+1)=1$ with
probability $A_i(T) \nu$. If $i$ is activated, then $S_i(T+1)=1$;
otherwise $i$ still needs to consider the following two cases:

\begin{enumerate}
\item  If $i$ has no history of being active then $S_i(T+1)=0$.

\item If $i$ has the history of being active, then he needs to recall the
most recent time $T^i_{1,0}$ when his state was switched from 1 to
0. He can remember the message with probability
$P_{+}(T-T^i_{1,0})$. If he successfully remembers the message, then
$S_i(T+1)=1$; otherwise $S_i(T+1)=0$.
\end{enumerate}

\end{enumerate}
\noindent

As we can see that the dynamics of the system is interesting and
very complex, mainly due to the existence of the FRM. The system is
mainly driven by the competition between the forget- and remember
mechanism. If the former prevails, there is great chance that the
message may die out at some final moment. On the contrary, the
message can be further spread.

\section{Results and discussions}

%\begin{figure}
%\includegraphics{fig2.eps}
%\includegraphics[width=18cm]
%\caption{$D(T)$, the percentage of active individuals, versus system
%time step $T$, with transmission probability $\nu=0.002$. The system
%size is 10,000. The solid curves correspond to the simulations with
%fixed remember probability $P_{+}=0.1$, while the dashed ones
%display those without remember effects. Here, the forget function
%takes the exponential form with $\alpha=1$ and $\beta=$ 0.001,
%0.005, 0.01, 0.03 and 0.05 (from top to bottom).} \label{fig.2}
%\end{figure}

%\begin{figure}
%\includegraphics{fig3.eps}
%\includegraphics[width=18cm]
%\caption{The stationary values of $D(T)$, calculated from the
%curves in Fig. 2, versus $\beta$. The round data points represent
%$D_{\rm s1}$ for the case with the remember mechanism, and the
%square ones represent $D_{\rm s2}$ for the one without.}
%\label{fig.3}
%\end{figure}

%\begin{figure}
%\includegraphics{fig4.eps}
%\includegraphics[width=18cm]
%\caption{$D(T)$ versus system time step $T$ with remember function
%taking the linear form with $b=0.001$ and $a=$ 0.01, 0.05, 0.1, 0.5
%or 1 (from bottom to top), forget probability $P_{-}=0.1$ and
%transmission probability $\nu=0.002$. The system size is 10,000.}
%\label{fig.4}
%\end{figure}

The effects of the above forms of forget- and remember functions
on spreading can be well demonstrated by computing a quantity,
$D(T)$, the percentage of active individuals. As shown in Figs. 2
and 4, the dependence of $D(T)$ on $T$ is sensitive to parameters
$a$, $b$, $\alpha$ and $\beta$. We have actually run numerous
simulations by spanning the wide range of a large parameter space,
which more or less display similar trends to those given by Figs.
2 and 4 \cite{Refz38}. First we come to Fig. 2, which corresponds
to the case where the forget probability can vary exponentially
and the remember probability is fixed to be 0.1. The below are the
observations for this part: (i) Generally $D(T)$ will saturate or
reach a stationary value after a certain number of time steps,
say, around 2000. This indicates that the convergence to the
stationary states is quick. (ii) The stationary value of $D(T)$,
denoted by $D_{\rm s}$, for the case without the remember
mechanism, $D_{s2}$, is much smaller than its counterpart with,
$D_{s1}$. This means that the existence of the remember mechanism
will be an advantage for message spreading, which is obvious. But
quantitatively we know how large the difference, namely $\Delta
D_{\rm s}=D_{\rm s1}-D_{\rm s2}$, is. For example, when $\alpha=1$
and $\beta=0.001$, the difference is around 0.2. For $\alpha=1$
and $\beta=0.01$, the difference is 0.72. More simulations show
the dependence of $\Delta D_{\rm s}$ on $\beta$, which is a curve
of first rapid increase and then slow variation. It can be seen
from Fig. 3 when the remember mechanism is included, $D_{\rm s1}$
decreases almost linearly with $\beta$; otherwise $D_{\rm s2}$
decreases exponentially with $\beta$. (iii) The saturation time
$T_{\rm c}$, namely the time step when $\partial D(T_{\rm
c})/\partial T_{\rm c}=0$, is nearly independent of parameter
$\beta$. (iV) When $\alpha=1$ and $\beta$ is as large as 0.05, the
spreading of the message comes to a halt at a very early stage.

The simulations of the situation in which the remember probability
varies and the forget probability is fixed to be 0.1 are given in
Fig. 4. Here are some observations: (i) The increase of $a$, with
fixed $b$, will accelerate the message spreading, which eventually
results in the increase of $D_{\rm s}$. For example, $D_{\rm s}$
is 0.1, 0.3, 0.5, 0.82 and 0.9 for $a=0.01$, 0.05, 0.1, 0.5 and 1,
respectively. This indicates that parameter $a$ plays a positive
role in the message spreading. (ii) There are certain cases where
the message can still be spread but not as effectively as in other
cases. For example: when $a=0.01$ and $b=0.001$, $D_{\rm s}$ is as
low as $0.1$, namely, 10 percent of the population is infected.
But we shall keep in mind that this percentage is still
considerable when we are dealing with an epidemic.

Let us now analyze how the above observations may provide hints to
help prevent an epidemic from breaking out. As we notice that if
there is no remember mechanism and the forget probability function
is $1-e^{-0.05t}$, the spreading will be terminated at the very
beginning. First, this means that vaccination (of course for
vaccinable diseases) is important. By vaccination you can greatly
reduce the ``remember mechanism", which protects you from being
``infected" by that same disease. The forget speed 0.05 then
suggests it is better to cure a disease as quickly as possible.
Otherwise, the prolongation of its cure duration may enhance the
risk of spreading it to others. Second, some diseases like flu may
not be vaccinable so can be re-infected. And its cure duration may
usually take a while such as weeks, which can be corresponding to
the fixed forget probability in Fig. 2. Therefore it is wise to get
less contact with patients. It would not be difficult to understand
that quarantine of the active hubs (the ill person who have many
acquaintances) may be an efficient way to prevent an epidemic.

In our model the stationary state, where $D(T)$ remains nearly
constant, can always be reached however we vary the parameters. Fig.
5 exhibits the relationship between $D_{\rm s}$ and both parameters
$a$ and $b$, where the simulations were performed under the
condition of $P_{-}(t)=0.1$, $P_{+}(t)=a-bt$ and $\nu=0.002$. The
value of $D_{\rm s}$ first increases with $a$ very rapidly and then
is stabilized. $D_{\rm s}$ can span the whole range between 0 and 1
as parameters are varied. That is to say, the FRM injects
significant effects into the spreading process.

To display more clearly the tendency of $D_{\rm s}$ versus both
parameters $a$ and $b$, we chose one special case among the results
shown in Fig. 5, where $b$ is 0.001 and $a$ can be varied. Fig. 6
clearly implies a transition at a certain value of $a_{\rm c}$,
below which there is null activity and above which the spreading
persists.

It can be inferred that the system can also switch from a more
complex phase to a simpler one. In the more complex phase,
individuals, no matter active or inactive, are scattered and
intermingled. In the simpler phase, nearly all individuals are
active and form a very huge cluster. We will now briefly explain why
such a switch can occur. As already stated in the previous section,
the spreading of a message or not is now mainly determined by the
competition between the forget mechanism and the remember mechanism
plus the transmission probability $\nu$. When the former mechanism
overcomes the sum of the latter two, the spreading can be terminated
or is at least not that efficient. On the contrary the message will
be spread to far away. Take the linear forget-remember function as
an example, when $a=1$ one will definitely remember a message before
the remember probability decays. The forget probability is always
0.1. Henceforth at $a=1$, the effect caused by the remember
mechanism plus the transmission probability is stronger than the one
caused by the forget mechanism alone, due to which the spreading is
efficient.

We also investigated the influence of system size $N$ on $D_{\rm
s}$, other conditions being equal. The results show that the
effect of the system size is not significant especially as $N$
grows. As indicated in Fig. 7, under the conditions of
$P_{-}(t)=0.1$, $P_{+}(t)=a-0.001t$ and $\nu=0.002$, the
three-dimensional figure (Fig. 7) displays the variation of
$D_{\rm s}$ versus both parameters $a$ (from 0 to 1.4) and $N$
(from 1000 to 10000). When $N$ is fixed, $D_{\rm s}$ increases
quickly (from 0 to 0.9) with $a$'s increasing (from 0 to 1). When
$a$ equals 1, $D_{\rm s}$ reaches the saturated value and is
nearly constant when $a>1$. Fig. 7 (a) also shows that $D_{\rm s}$
does not change drastically with $N$ when $a$ is fixed. This
phenomenon is more clearly demonstrated in Fig. 7 (b), where the
variation of $D_{\rm s}$ versus $N$ was given, when $a=$0.2, 0.4,
0.6, 0.8 and 1 (from bottom to the top). We find that $D_{\rm s}$
for smaller $N$ ($N \le 3000$) is slightly smaller than its
counterpart for larger $N$ ($N > 3000$). But this difference
becomes vanishing as $N$ grows. For example, when $N
>3000$, $D_{\rm s}$ maintains a steady value of 0.8 for $a=0.4$. The
system size $N$ of our major simulations is 10000, so the
finite-size effect is almost negligible.

We performed ensemble analysis of our model via simulations of
different network realizations. We found as $N$ is large enough,
there is no significant difference between the outcomes of different
realizations. For accuracy, our major results were averaged over 10
different network realizations.

In the message spreading, it is very obvious that the transmission
probability $\nu$ has a significant effect on $D_{\rm s}$. Fig. 8
shows the variation of $D_{\rm s}$ with $\nu$ through simulations
under the conditions of $P_{-}(t)=1-e^{-0.03t}$, $P_{+}(t)=0.1$
(top curve)and $P_{-}(t)=0.1$, $P_{+}(t)=0.01-0.001t$ (bottom
curve). The larger $\nu$ is, the larger $D_{\rm s}$ will be. For
example, $D_{\rm s}$ is 0.5 for $\nu=0.05$, and 0.7 for $\nu=0.2$,
for the bottom curve. Hence the choice of $\nu$ should be careful,
for if it is too small (close to zero), it will restrict the
message spreading. But if it is too large the message spreads too
quickly and the effects of the FRM will be covered. In most of our
simulations $\nu$ is set to be 0.002.

The influence of initial conditions $P_{\rm A}$, defined as the
percentage of initially activated nodes, was also considered in
our simulations. The results show that $P_{\rm A}$ has no
significant influence on the message spreading provided $P_{\rm
A}$ is not within the regime adjacent to zero. For example, with
$\nu=0.002$, $P_{-}=0.1$ and $P_{+}(t)=a-0.001t$ (Fig. 9), $D_{\rm
s}$ increases rapidly as $P_{\rm A}$ does from 0 to 0.001.
However, as long as $P_{\rm A}$ is chosen to be larger than 0.001,
$D_{\rm s}$ is nearly independent of $P_{\rm A}$. In our
simulations, we chose $P_{\rm A}$ to be 0.005.

The time scales of the transitions between inactive and active
states are the parameters determining the behavior of spreading. We
define $T_{\rm c}$ the time for $D(T)$ reaching the saturated value
$D_{\rm s}$ for the first time. Take the case of $\nu=0.002$,
$P_{-}=0.1$ and $P_{+}(t)=a-0.001t$ as an example (Fig. 10), there
exists power-law relationship between $T_{\rm c}$ and $a$. It was
also found that the exponent of such a power-law is -0.2, nearly
independent of system sizes.

We hereby show the simulations when neither remember- nor forget
probability is constant. Namely, we shall pay attention to the
cases in which all the four parameters $a$, $b$, $\alpha$ and
$\beta$ are non-zero variables, despite that the situation is very
complicated. Fig. 11 displayed the results of 8 sets of different
combinations, including the one where both functions are linear
(e.g. $P_{-}(t)=0.5+0.05t$ and $P_{+}(t)=0.5-0.05t$), the one
where both are exponential ( e.g. $P_{-}(t)=0.05-e^{-0.05t}$
and$P_{+}(t)=0.05+e^{-0.05t}$), the one where one is linear and
another is exponential (e.g. $P_{+}(t)=0.5-0.05t$ and
$P_{-}(t)=0.05-e^{-0.05t}$), and the one where the parameters are
same (e.g. $P_{-}(t)=0.05-e^{-0.05t}$
and$P_{+}(t)=0.05+e^{-0.05t}$) and the one where the parameters
are distinct (e.g. $P_{-}(t)=0.5+0.05t$ and $P_{+}(t)=0.5-0.05t$).
These simulations, nevertheless, show rather similar trends to the
ones already observed in Fig. 2. More detailed analysis concerning
this part will be provided in our next paper.

In order to characterize the phase transition mentioned above, we
consider the clusters that are formed by connected active
individuals. The size of a certain cluster is the number of active
individuals within it. The first quantity of interest is the size
distribution of clusters. We see that in Fig. 12 (a), at small value
of $a$, only small, isolated clusters can be formed. As shown in
Fig. 12 (b), "infinite" clusters start to form only after a certain
value of $a$. This observation is very similar to the percolation,
where below the critical density $P_{\rm c}$ only small clusters can
be formed and above $P_{\rm c}$ larger clusters with sizes
comparable to the system size come into being. Fig. 13s (a) and (b)
display the variation of the size distributions of clusters with the
parameter $a$. These two panels equivalently exhibit the transition
displayed by Fig. 12s (a) and (b).

The second quantity of interest is the size of the largest cluster
ever formed, denoted by $S_{\rm max}$. Fig. 14 shows the relation
between $S_{\rm max}$ and $a$, where the data of 5000 steps were
taken after the system reaches a stationary state. We notice that
the value of $S_{\rm max}$ increases with the increasing parameter
$a$ and becomes stable later on. We note in Fig. 14 that the initial
variation of $S_{\rm max}$ is very steep, which can be fitted by an
exponential function. Correspondingly, the average size of clusters
(excluding the largest cluster) versus $a$ is now given by Fig. 15.
We note that initially $\langle S \rangle$ is almost a constant,
close to 1, with respect to $a$. After $a$ surpasses the point
$a=0.01$, $\langle S \rangle$ starts to increase and finally
diverges exponentially (well fitted to 21.6$e^{2.28a}$). This may
imply that the critical value of $a$ is small for the parameters
used, which needs scrutiny through more systematic analysis in the
future work.

In the previous study of the spreading process, the epidemic
processes in an uncorrelated network possess an epidemic threshold
on the scale-free network, below which the diseases cannot produce a
macroscopic epidemic outbreak \cite{Refz39,Refz40}. Correspondingly
in our model, the value of $D(T)$ is determined by the parameters of
the probability functions when one fixes the transmission
probability $\nu$. There also exists certain threshold in the FRM.
For example, when $a$ is as small as 0.001, the message can not be
spread effectively.

\section{Conclusion}

In summary, we have presented a simple forget-remember mechanism for
studying the spreading process. We have investigated how the FRM
affects the spreading when we vary the parameters of the forget- and
the remember, probability function. Our main results are: (i) When
the transmission probability is vanishingly small, the competition
between the forget- and the remember, mechanism is the main force to
drive the system to the stationary state. When the forget effect
prevails, the spreading may not be efficient mostly. (ii) When there
exists remember mechanism in the system, there is great chance for
an ``epidemic" to form. When the remember effect is none or weak,
the message may be spread less effectively than it does with a
stronger remember effect. Hence this suggests that by vaccination or
having less contact with the infected individuals may protect one
from being infected. (iii) There is a phase transition that can be
characterized by the divergence of the average size of clusters,
formed by active individuals, in the critical regime. (iv) The
outcome of our model is sensitive neither to the system size as long
as it is large enough ($>3000$), nor to the initial condition (the
percentage of initially activate nodes). But the outcome is
sensitive to the transmission probability which may cover the
effects of the forget-remember at the larger values. This indicates
that the forget-remember mechanism dominates the transmission
probability only when the latter is small enough.

The FRM on the spreading system can help to explain why different
diseases have different saturation values in a population, and we
hope this mechanism can be well applied to solving practical
problems. For example, if immunity in humans is enhanced, the
initial probability of relapse decreases. People who know they may
be exposed to a specific disease can get medication and otherwise
prepare for it, increasing the probability that if they become ill,
they will recover, and reducing the probability of relapse, which
corresponds to adjusting the forget- and remember probability in our
model. This might prevent the occurrence of diseases on a large
scale.

\section{Acknowledgments} The authors would like
to thank Laura Ware of the Santa Fe Institute for type-editing of
the manuscript. W.L. would like to thank Professor Jost of
Max-Planck-Institute for Mathematics in the Sciences for hospitality
during his stay at the institute where part of this work was done.
This work was in part supported by the National Natural Science
Foundation of China (Grant Nos. 70571027, 10647125, 10635020 and
70401020) and the Ministry of Education of China (Grant No. 306022
and the "111" project with Grant No. B08033).

\end{document}